\documentclass[twocolumn,showpacs,preprintnumbers,amsmath,amssymb]{revtex4}
\usepackage{graphics}
\usepackage{epsfig}
\usepackage{color}
\usepackage{subfigure}

\begin{document}

\title{Can one measure negative specific heat in the canonical statistical ensemble?}
\author{F. Staniscia$^{1,2}$, A. Turchi$^{3}$, D. Fanelli$^{4}$ P.H. Chavanis$^{5,6}$, G. De Ninno$^{7,2}$}
\affiliation{1. Dipartimento di Fisica, Universit\`a di Trieste, Italy \\
2. Sincrotrone Trieste, S.S. 14 km 163.5, Basovizza (Ts), Italy \\
3. Dipartimento di Sistemi e Informatica, Universit\`a di Firenze, via S. Marta 3, 50139 Firenze, Italy\\
4. Dipartimento di Energetica ``Sergio Stecco'', Universit\`a di Firenze, via S. Marta 3, 50139 Firenze, Italy\\
5. Universit\'e de Toulouse, UPS, Laboratoire de Physique Th\'eorique (IRSAMC), F-31062 Toulouse, France\\
6. CNRS, Laboratoire de Physique Th\'eorique (IRSAMC), F-31062 Toulouse, France\\
7. Physics Department, Nova Gorica University, Nova Gorica, Slovenia
}

\date{\today}

\begin{abstract}
According to thermodynamics, the specific heat of Boltzmannian short-range interacting systems is a positive quantity. Less intuitive 
properties are instead displayed by systems characterized by long-range interactions. In that case, the sign of specific heat depends on the 
considered statistical ensemble: negative specific heat can be found in isolated systems, which are studied in the framework of the 
microcanonical ensemble; on the other hand, it is generally recognized that a positive specific heat should  always be measured in  
systems in contact with a thermal bath, for which the canonical ensemble is the appropriate one. We demonstrate that the latter assumption is not generally true: 
one can in principle measure negative specific heat also in the canonical ensemble if the system under scrutiny is non-Boltzmannian and/or out-of-equilibrium.  
\end{abstract}

\maketitle

In classical thermodynamics, specific heat is defined as the energy required to $\it{increase}$ the temperature of a unit quantity
of a substance by a unit of temperature. Such a definition follows common sense:
providing energy to a physical system should induce heating, that is an increase of the system temperature.
In fact, this is normally the case of physical systems obeying the Boltzmann statistics
\cite{Huang} and being characterized by ``short-range'' interactions. In a short-range interacting system, the interaction potential 
between different components decays as $1/r^\alpha$ with $\alpha > d$ ($r$ being the distance
and $d$ the system dimension). However, less intuitive thermodynamic properties may be displayed
by ``long-range'' interacting (LRI)
systems, for which $\alpha < d$. In that case, experiments realized on isolated systems, described by
the micocanonical statistical ensemble, may give different results from similar experiments performed on systems in contact with a thermal bath, for which the canonical ensemble is the appropriate one. According to literature (see, e.g., \cite{revSR} and references therein), one
of the most striking features of such ensemble inequivalence is the presence of a negative specific heat in
the microcanonical ensemble: increasing the energy of an isolated LRI system may lead to a decrease of its thermodynamic
temperature. On the other hand, it is generally recognized that the specific heat in the canonical ensemble
is always positive, even if interactions are long range. Hence, it is a generally shared opinion that one can not $\it{measure}$ negative 
specific heat in a system in contact with a thermal bath. As we will see, such a result relies on the assumption that
the thermodynamic temperature, defined as $T_\text{th}=(\partial s/\partial e)^{-1}$ (where $s$ and $e$ are, respectively, the system entropy 
and the energy per particle), coincides with the kinetic temperature, $T_\text{kin}$, defined in terms of the average of the system kinetic energy. In fact, this is true
only for Boltzmannian systems, and generally not for systems following a different statistics. 

In this Letter, we demonstrate that negative specific heat in the statistical canonical ensemble can be measured in
non-Boltzmannian and/or out-of-equilibrium systems. Moreover, the common opinion according to which measuring a negative specific heat (in the
microcanonical ensemble) is to be considered as a signature of ensemble inequivalence should be revised: we will show that
ensemble equivalence may still hold when negative specific heat is measured (in both ensembles). 

As a paradigmatic example for our investigation, we consider the so-called Hamiltonian Mean Field (HMF) model \cite{hmf}, which has been
widely studied in the past as a prototype of LRI system. The HMF model, which shares many similarities with gravitational and charged sheet
models, describes the one-dimensional motion of $N$ rotators coupled 
through a mean field cosine interaction. The system Hamiltonian reads
\begin{equation}
\label{hamil}
H = \frac{1}{2} \sum_{j =1}^{N} p_{j}^2 + \frac{1}{2N}
\sum_{i,j=1}^{N} [ 1-\cos ( \theta_j - \theta_i)]
\end{equation}
where $\theta_j$ represents the orientation of the $j$-th rotator and $p_j$ stands for its conjugated momentum. To monitor the 
system evolution, it is customary to introduce the magnetization $M$, an order parameter defined as
\begin{equation}
M = \frac{\left||\sum_i \mathbf{m}_i\right||}{N} \quad \mbox{where}
\quad \mathbf{m}_i = ( \cos \theta_i , \sin \theta_i) \label{l}.
\end{equation}

The infinite-range coupling between rotators, provides the system dynamics with all typical characteristics of a LRI system. In 
particular, it is well known that after a transient regime (the so-called ``violent relaxation''), and before
attaining the final Boltzmann equilibrium, the system may be trapped in a ``quasi-stationary state'' (QSS), whose lifetime
diverges with the system size $N$ \cite{hmf, latora}. In this regime, the magnetization is lower than predicted by the
Boltzmann equilibrium and the system displays non Gaussian velocity distributions \cite{nonGauss}. It has been shown that QSS's can be related to the stable steady states of the Vlasov equation describing the system in the limit
$N\rightarrow \infty$ \cite{yama}. The idea was inspired by the seminal work of Lynden-Bell \cite{Lynden-Bell}, developed in the context of stellar dynamics, and later applied
to vortex dynamics (see, e.g., \cite{chava1}). 

Lynden-Bell's approach goes as follows. Starting from an arbitrary initial condition, the system evolution makes the single-particle distribution function 
stir in phase space. However, as a property of the Vlasov equation, the hypervolumes of its values $\eta$ (level of phase density), are conserved. Let us introduce the probability density $\rho(\theta,p,\eta)$ 
of finding the levels of phase density $\eta$ in a small neighborhood of the position $(\theta, p)$ in phase space. Using $\rho$, 
Lynden-Bell defined a locally averaged 
(``coarse-grained'') distribution function $\bar{f}(\theta,p)=\int \rho(\theta,p,\eta) \eta d \eta$, and, in order to determine the system 
equilibrium distribution, introduced an entropy functional, like in ordinary statistical mechanics: $s=-\int \rho(\theta,p,\eta) \ln \rho(\theta,p,\eta) d \eta d \theta d p$. Following Lynden-Bell, 
we consider the particular case of two-level water-bag initial conditions, i.e. $\eta=f_0$ or  $\eta=0$. Within this approximation, the entropy reduces to

\begin{equation}
s[\bar{f}]=-\int \!\!{\mathrm d}p{\mathrm d}\theta \,
\left[\frac{\bar{f}}{f_0} \ln \frac{\bar{f}}{f_0}
+\left(1-\frac{\bar{f}}{f_0}\right)\ln
\left(1-\frac{\bar{f}}{f_0}\right)\right],
\label{eq:entropieVlasov}
\end{equation}
where $\bar{f}=f_0 \rho$. As we will show in the following, maximizing this entropy, one obtains a ``fermionic'' solution, $\bar{f}_{\text{QSS}}(\theta,p)$,
which represents the most probable distribution when the system is trapped in the QSS, and which coincides with a particular class of stable 
stationary solutions of the Vlasov equation. While QSS's represent out-of-equilibrium states of the $N$-body dynamics, they could be equally interpreted as equilibrium configuration of the corresponding continuous description: in this respect, the conclusions of our analysis will apply to both equilibrium and non equilibrium dynamics, provided the latter bears distinctive non-Boltzmannian traits.

Before proceeding further, it is worth emphasizing that our study is relevant for a quite large class of systems and for many practical physical situations. 
In fact, experimental LRI systems often involve a very large number of components. When this is the case, QSS's may last long enough to prevent the system from reaching 
the final Boltzmann equilibrium during the finite time of an experimental session. For all these experiments, QSS's represent the effective (measurable) equilibrium states. 
As an example, one can mention charged systems, e.g. plasmas \cite{dubin}, systems involving wave-(many) particle interactions, e.g. free-electron lasers \cite{fel} 
and collective atomic recoil lasing (CARL) \cite{carl}, and magnetic dipolar systems \cite{dip}. Long lasting QSS's may be also displayed by 
gravitational systems (e.g., galaxies), 
to which Lynden-Bell theory was first applied. The model we are here considering, although very simple, reproduces the most relevant features of the above mentioned 
systems. In particular, it can be shown that, under some hypotheses, the Hamiltonian (\ref{hamil}) can be formally reduced to that governing CARL dynamics \cite{rom}. This important observation will make it eventually possible to test our prediction versus direct experiments. 
Further, it is important to stress that for systems characterized by small energy dispersions, the two-level water-bag initial conditions represent a 
good approximation of a more natural Gaussian initial distribution \cite{fel}. 

In the following, we briefly review the standard argument leading to the conclusion that the specific heat in the canonical ensemble is always positive, and we discuss in which respects such an argument does not apply to our case. 

Consider a generic (long-range or short-range) system. Given the microcanonical entropy $s(e,n)$, where $n$ is the particle density, the canonical rescaled free energy, $\phi(\beta,n)=\beta f(\beta,n)$, can be calculated as the 
Legendre-Fenchel Transform (LFT) of $s(e,n)$: $\phi(\beta,n)=\mbox{inf}_{e}\left[\beta e -s(e,n)\right]$. Here $\beta$ is a free parameter to be related to the system temperature.
It is worth reminding that the LFT of a generic function is always a concave function. If $s(e,n)$ is also concave, the inverse LFT can be
applied: $s(e,n)=\mbox{inf}_{\beta}\left[\beta e - \phi(\beta,n)\right]$. This is always the case of short-range systems, for which the equivalence of statistical ensembles holds true. Using the previous relation one can write
\begin{equation}
\frac{\partial^2 s}{\partial e^2} = \frac{\partial \beta}{\partial e} = \frac{\partial \beta}{\partial T} \frac{1}{ c_v^\text{mic}},
\label{dersec}
\end{equation}
where 
\begin{equation}
c_v^\text{mic}=\frac{\partial e}{\partial T}=\frac{1}{\displaystyle{\left(\frac{\partial^2 s}{\partial e^2}\right)}\left(\frac{\partial T}{\partial \beta}\right)}
\label{calo1}
\end{equation}
is the specific heat (at fixed volume) calculated in the microcanonical ensemble and $T$ is either the thermodynamic ($T_\text{th}$) or the kinetic ($T_\text{kin}$) 
system temperature.  

Making use of the LFT, the specific heat can be also calculated in the canonical ensemble and expressed in terms of the rescaled free energy: 
\begin{equation}
c_v^\text{can}=\frac{\displaystyle{\left(\frac{\partial^2 \phi}{\partial \beta^2}\right)}}{\displaystyle{\left(\frac{\partial T}{\partial \beta}\right)}}.
\label{calo2}
\end{equation}

Using the thermodynamical definition of temperature: $T_\text{th}=(\partial s/\partial e)^{-1}=1/\beta$, one gets $\left(c_v^\text{mic}\right)_\text{th}=-1/T_\text{th}^2\left(\partial^2 s/\partial e^2\right)$. 
The sign of $\left(c_v^\text{mic}\right)_\text{th}$ depends on that of $\left(\partial^2 s/\partial e^2\right)$: for concave entropies (short-range systems), 
i.e. $\partial^2 s/\partial e^2 <0$, $\left(c_v^\text{mic}\right)_\text{th} > 0 $; if the entropy has a ``convex intruder'', 
i.e. $\partial^2 s/\partial e^2>0$ in some energy range, $\left(c_v^\text{mic}\right)_\text{th} <0 $. The latter may be the case of LRI systems 
in the presence of a first-order phase transition \cite{revSR}. In the canonical ensemble, 
$\left(c_v^\text{can}\right)_\text{th}=-\frac{1}{T_\text{th}^2} \left(\frac{\partial^2 \phi}{\partial \beta^2}\right)$. As we have already mentioned, 
$\phi$ is always concave and, as a consequence, $\left(c_v^\text{can}\right)_\text{th}$ is always positive. 

Consider now the definition of kinetic temperature, which is the one experimentally accessible \cite{bale}:  
\begin{equation}
T_\text{kin}=\int p^2 F(\theta,p) d \theta d p
\label{temp}
\end{equation}
(we assume unitary mass in the definition of kinetic energy).
Here $F(\theta,p)$ is the normalized distribution function of 
the system. Eq. (\ref{temp}) furnishes the link between $T_\text{kin}$ 
and the thermodynamical temperature $T_\text{th}=1/\beta$. 
For a Boltzmannian system: $F(\theta,p)=C \exp(-\beta p^2/2)$, where $C$ is a normalization constant \cite{Huang}. 
Substituting the latter expression into Eq. (\ref{temp}) one gets $T_\text{kin}=1/\beta= T_\text{th}$. In this case the statistical 
and kinetic definitions of specific heat provide the same result: $\left(c_v\right)_\text{th}=\left(c_v\right)_\text{kin}$.     

From the above, one can conclude that, for a Boltzmannian system, the canonical specific heat is
always positive, regardless the definition of temperature. As a consequence, for a Boltzmannian LRI system, the presence of negative 
specific heat in the microcanonical ensemble is the signature of ensemble inequivalence. 

A completely different scenario may arise in the case of non-Boltzmannian or out-of-equilibrium LRI systems. As we shall demonstrate, for those systems one can expect to measure negative specific heat in the canonical ensemble. 

Consider the ``fermionic'' system characterized by the entropy (\ref{eq:entropieVlasov}), paradigmatic example of a LRI system ``trapped'' in a QSS. Following the procedure outlined above, one can perform the LFT of $s[\bar{f}]$ and calculate the rescaled free energy $\phi[\bar{f},\beta]$. Requiring the free energy to be stationary, one gets the following distribution:   

\begin{multline}
  \label{eq:barf}
  \bar{f}_{\text{QSS}}(\theta,p)= \\
  \frac{f_0}{ 1+e^{\displaystyle\beta f_0 (p^2/2 -M_x[\bar{f}_{\text{QSS}}]\cos\theta-M_y[\bar{f}_{\text{QSS}}]\sin\theta)+\alpha}}.
\end{multline} 
Here $\alpha$ plays the role of a Lagrange multiplier associated, rispectively, to mass conservation, while $M_x[\bar{f}]=\int \bar{f} \cos(\theta) d\theta d p$, $M_y[\bar{f}]=\int \bar{f} \sin(\theta) d\theta d p$ stand for the two components of magnetization in the $N\rightarrow \infty$ limit. 
It is worth stressing that the expression (\ref{eq:barf}) for the stationary distribution in the canonical ensemble is formally identical to the 
one obtained by requiring the entropy (\ref{eq:entropieVlasov}) to be stationary in the framework of the microcanonical ensemble \cite{nonGauss}. 
In that case, $\beta$ and $\alpha$ are Lagrange multipliers associated to energy and mass conservation.    

\begin{figure}[t]
\resizebox{0.48\textwidth}{!}{\includegraphics{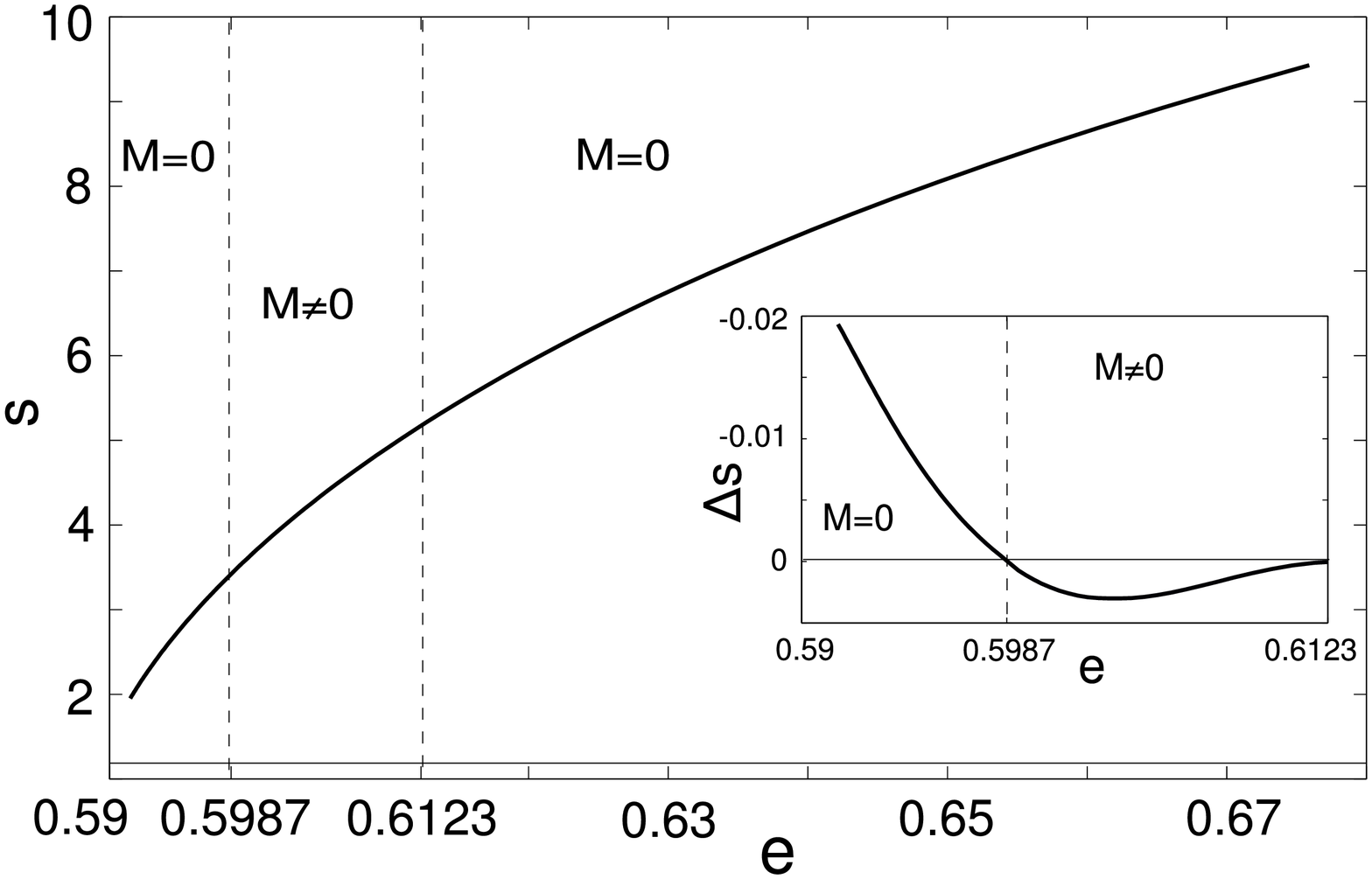}}
\caption{Lynden Bell entropic curve, $s=s(e)$, calculated for the initial condition $f_0=1.1096$, in the framework of the microcanonical ensemble. 
Inset: Difference, $\Delta s$, between magnetized and non-magnetized branches of the entropic curve, as a function of energy. A change of sign 
in $\Delta s (e)$ is the signal of a phase transition, see also \cite{fabio}. There is a second-order phase transition at $e=0.6123$ and a
first-order phase transition at $e=0.5987$. Note that the latter cannot be noticed on the (too large) scale used to plot the entropic curve. \label{fig1}}
\end{figure}

Fig. \ref{fig1} shows the entropic curve, $s=s(e)$, for a given initial condition $f_0$. The curve is calculated in the framework of the microcanonical ensemble, using the following procedure. 
Once fixed $e$ and $f_0$, one first determines the corresponding values of $M_x$, $M_y$, $\alpha$ and $\beta$ in the expression (\ref{eq:barf})
for $\bar{f}_\text{QSS}$. This can be done by imposing energy and mass conservation and by using the definition of $M_x$ and $M_y$
\cite{nonGauss}. In general, one obtains several possible distributions $\bar{f}_\text{QSS}$, corresponding either to magnetized (i.e.,
$M=\left(M_x^2+M_y^2\right)^{1/2} \neq 0$) or to non-magnetized 
(i.e., $M = 0$) states. The value of $s(e)$ is found by substituting the possible solutions into (\ref{eq:entropieVlasov}) and by 
solving numerically the integral: the retained value is the one corresponding to maximum. This also selects the expected (magnetized 
or non-magnetized) stable equilibrium distribution. 

The Lynden-Bell entropic curve $s(e)$ is always concave (see Fig. \ref{fig1}): such a result is a clear indication of ensemble equivalence.  
Assuming $F(\theta,p)=\bar{f}_\text{QSS}(\theta, p)$ in Eq. (\ref{temp}), one gets a formal relation between the kinetic temperature, $T_{kin}$, 
and the Lynden-Bell Lagrange multiplier $\beta=1/T_\text{kin}$ (inverse thermodynamic temperature). 

Reconsidering now equations (\ref{calo1}) and (\ref{calo2}), with $T=T_\text{kin}$, one is led to the surprising conclusion that the measured specific heat can be negative both in the microcanonical and in the canonical ensembles, 
no matter the convexity of the entropy and of the free energy functions. 
In fact, in both cases the sign of the specific heat also depends on that of the derivative of the function $T_{kin}= T_{kin}(\beta)$. 

\begin{figure}
\centering
\subfigure
{\resizebox{0.48\textwidth}{!}{ \includegraphics{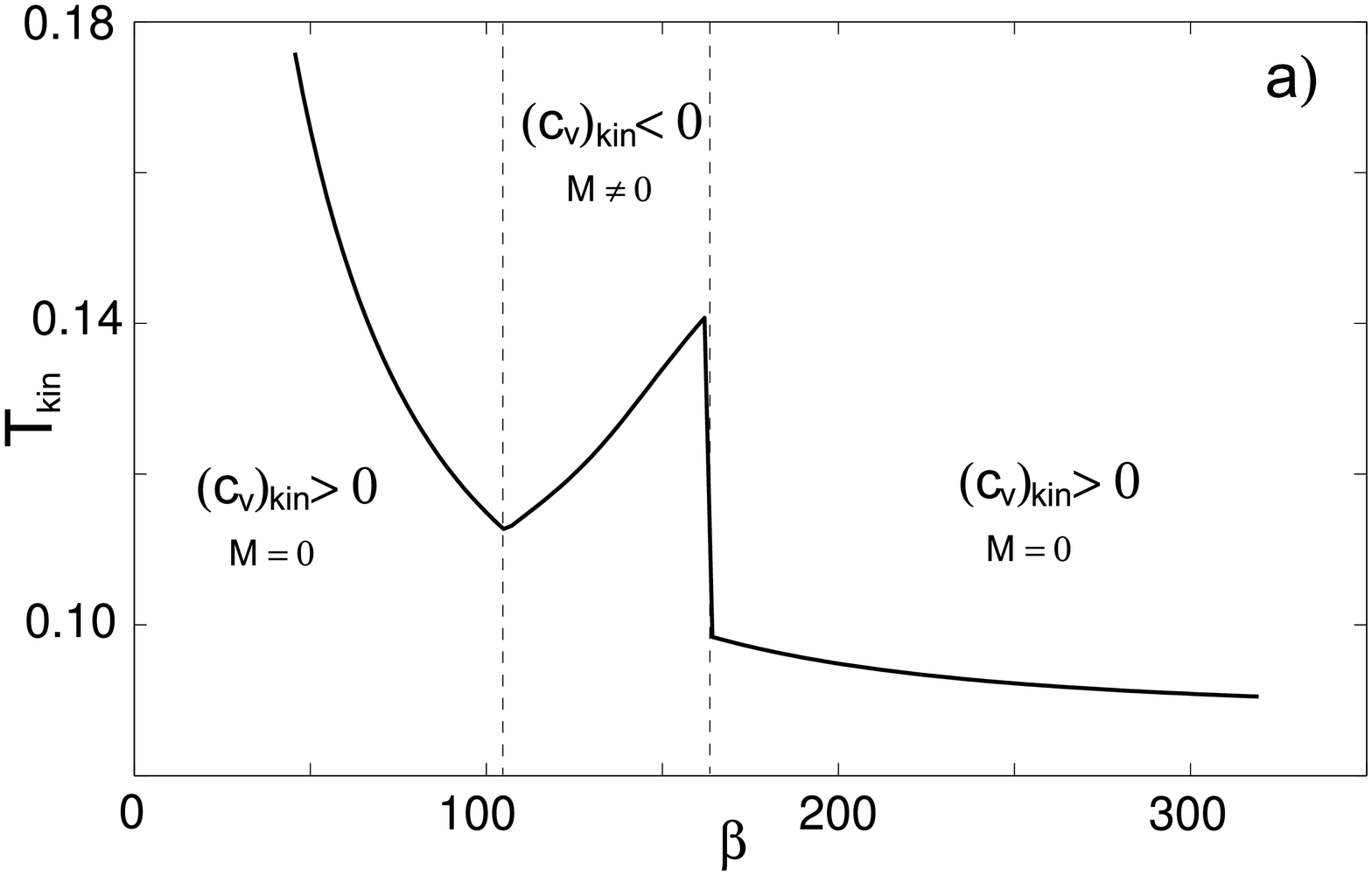}}}
\subfigure
{\resizebox{0.48\textwidth}{!}{ \includegraphics{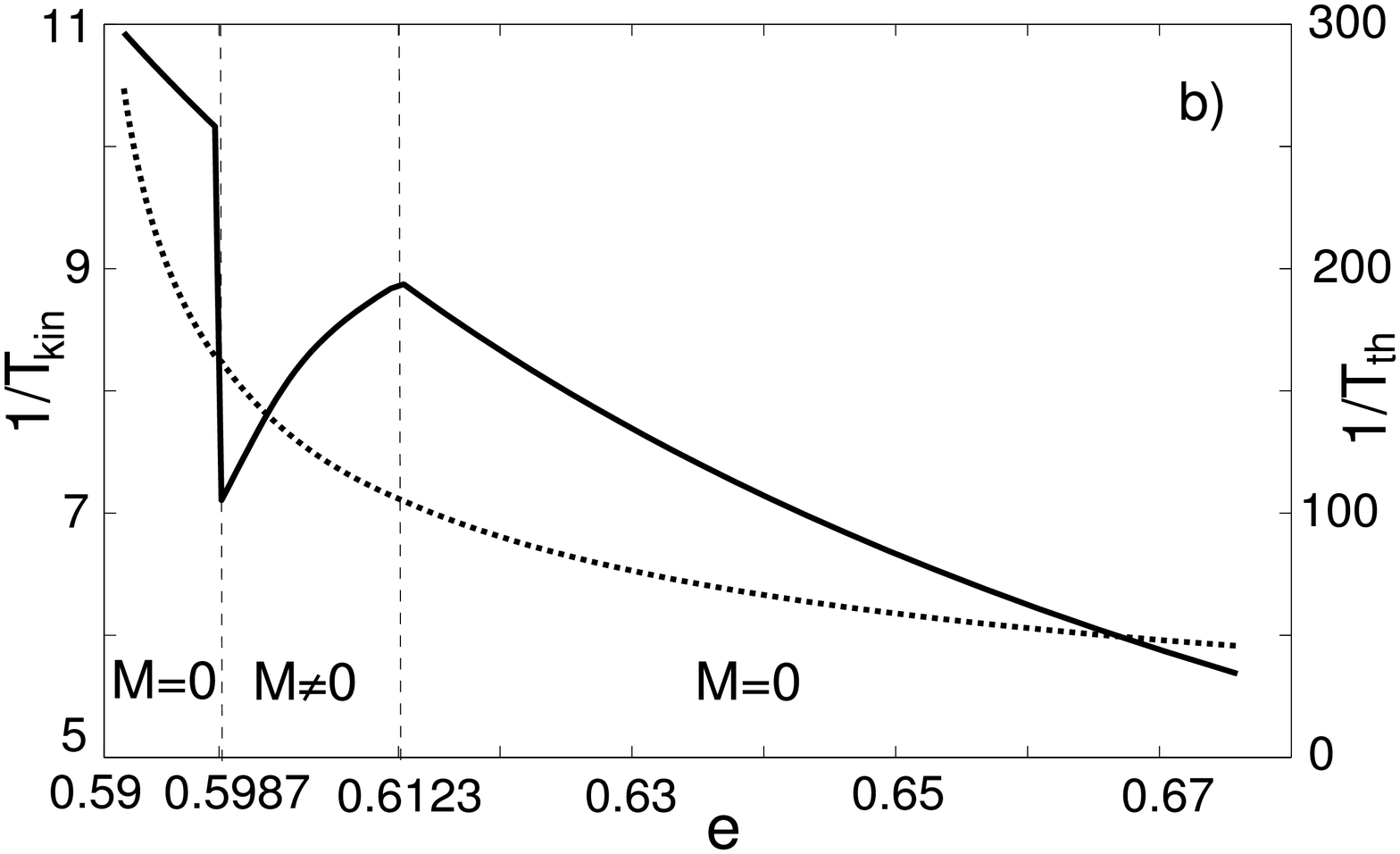}}}
   \caption{Fig. a): Kinetic temperature, $T_\text{kin}$, as a function of the Lynden-Bell Lagrange multiplier $\beta$, calculated for the initial condition $f_0=1.1096$. Fig. b): Kinetic caloric curve, $1/T_\text{kin}$ vs. $e$ (continuous line), and thermodynamic caloric curve, $1/T_\text{th}=\beta$ vs. $e$ (dotted line), calculated at $f_0=1.1096$.\label{fig2}}
\label{fig:averages}
\end{figure}

The curve $T_{kin}= T_{kin}(\beta)$ can be calculated in the framework of the canonical ensemble, as follows. Once fixed 
$\beta$ and $f_0$, one determines the stable equilibrium distribution by finding out the corresponding 
values of $M_x$, $M_y$ and $\alpha$. This can be done by imposing the normalization constraints and by using the definition of 
magnetization \footnote{Note that finding the equilibrium distribution in the framework of canonical ensemble 
requires to determine three parameters (and, thus, to solve three equations), instead of four, as required in the case of the microcanonical 
ensemble. This is due to the fact that energy does not explicitly enter the expression (\ref{eq:barf}) for the stationary distribution.}. The obtained result is 
then substituted into Eq. (\ref{temp}) and the integral solved numerically. The curve $T_\text{kin}= T_\text{kin}(\beta)$ is shown in Fig. \ref{fig2}a.

As it can be seen, one gets $\partial T_{kin}/(\partial \beta)>0$ and, thus negative specific heat, $\left(c_v \right)_\text{kin}<0$, in 
both ensembles, in the magnetized region. The change of sign occurs in correspondence of phase transitions: from positive to negative in 
correspondence of a second-order phase transition, at smaller $\beta$ values, and from negative to positive in correspondence of a 
first-order phase transition, at larger $\beta$ values, where the curve displays a discontinuity. Note that this behaviour is fundamentally different from that displayed by
Boltzmannian LRI systems. Indeed, as already mentioned, in that case the change from positive to negative (thermodynamical) specific heat may occur only in correspondence of a 
first-order phase transition \cite{ruffo2}.  
Phase transitions are 
also signaled by a change of stability of the magnetized and non-magnetized branches of the entropic curve, as shown in the inset of Fig. \ref{fig1}.
The presence of a negative kinetic specific heat in the magnetized region can be also noticed by looking at the kinetic caloric curve shown in Fig. \ref{fig2}b 
(continuous line). As expected on the basis of the result shown in Fig. \ref{fig1}, the thermodynamic caloric curve (dotted line in Fig. \ref{fig2}b), 
is instead characterized by a negative derivative over the whole energy range, corresponding to $\left(c_v \right)_\text{th} > 0 $ in both ensembles. 
It is worth stressing that the occurrence of negative specific heat in the microcanonical ensemble when the system is magnetized is confirmed by the 
results of direct $N$-body simulations, based on the Hamiltonian (\ref{hamil}) \cite{fabio}. 

As we anticipated, a further surprising result is that the presence of negative specific heat is associated to equivalent statistical ensembles. A
clear indication of ensemble equivalence can be obtained by inspection of Fig.  \ref{fig1}, showing a concave entropic curve, and of Fig. 
\ref{fig2}a, showing the occurrence of positive and negative canonical and microcanonical specific heats in the same interval of $\beta$ values.
The same indication can be drawn by the thermodynamical caloric curve shown in Fig. \ref{fig2}b. However, the curves displayed in figs. \ref{fig1}
and \ref{fig2} have been obtained for a particular initial condition $f_0$. In order to investigate ensemble equivalence for any $f_0$, we have
compared the values of $M$ obtained by solving the Lynden-Bell problem in the two ensembles for different initial conditions, following the 
above outlined procedures. Results (not reported here) show that the magnetized and unmagnetized regions perfectly coincide, and this confirms ensemble equivalence. 

Our study demonstrates the unexpected possibility to measure negative specific heat in the statistical canonical ensemble. As we argued above, this conclusion does not contradict any fundamental law of thermodynamics but ultimately originates from the non-Boltzmannian features that are possibly
associated to non-equilibrium, as well as equilibrium, dynamics. With reference to the case at hand, and for a specific window of the parameter space that we traced back to magnetized QSS,
the particles gain in kinetic energy when an energy quota is passed from the system to the heat reservoir. This surprising effect arises spontaneously and it is driven by the inherent ability of the system to
self-organize at the microscopic level. The average particle velocity is enhanced at the detriment of the potential energy, and this yields in turn to an increase of the experimentally measurable kinetic temperature. The opposite holds if the energy flows towards the system: the particles cool down, while the potential contribution grows so to guarantee for the needed energy balance. 
This phenomenon, that we have here demonstrated with reference to the
HMF model, is in principle general and can potentially extend to all those settings where non-Boltzmannian effects play a role. It is particularly attractive to speculate on the possibility of realizing
efficient thermal devices, working with a non-Boltzmannian fluid. Can one improve over current implementation by exploiting the aforementioned tendency to concentrate the residual energy amount into a kinetic quota that
supports and enhances the particle motion? As a final comment, let us emphasize that, by invoking the above mentioned analogy with the HMF model, dedicated CARL experiments should allow for a direct verification of the
predictions here elaborated and so contribute to shed light on the proposed scenario.

\end{document}